\documentclass[11pt]{article}
\usepackage[utf8]{inputenc}
\usepackage{authblk}
\usepackage{geometry}
\usepackage{amsmath}
\usepackage{graphicx}
\usepackage[numbers,sort&compress]{natbib}
\setcitestyle{numbers,square,sort&compress}
\usepackage{hyperref}
\makeatletter
\renewcommand\AB@affilsepx{\protect\\}
\makeatother

\title{\textbf{Machine Olfaction and Embedded AI Are Shaping the New Global Sensing Industry}}

\author[1,2,3]{Andreas Mershin}
\author[1]{Nikolas Stefanou}
\author[1]{Adan Rotteveel}
\author[1,3,4]{Matthew Kung}
\author[5,6]{George Kung}
\author[7]{Alexandru Dan}
\author[1,8]{Howard Kivell}
\author[1]{Zoia Okulova}
\author[1]{Zoi Kountouri}
\author[9]{Paul Pu Liang}

\affil[1]{RealNose.ai, 626 Massachusetts Ave., 2nd Floor, Arlington, MA 02476, USA}
\affil[2]{MIT Sloan School of Management, Massachusetts Institute of Technology, 77 Massachusetts Avenue, Cambridge, MA 02139, USA}
\affil[3]{The Osmocosm Public Benefit Foundation, www.OsmoCosm.org, Boston, MA, USA}
\affil[4]{Boston University, Boston, MA, USA}
\affil[5]{KungHo Fund, www.kungho.com, Cambridge, MA, USA}
\affil[6]{Harvard University, Cambridge, MA, USA}
\affil[7]{NATO DIANA \& TVL TECH Bdul. I. C. Bratianu 44 Bis, Bucharest, Romania}
\affil[8]{Endless Frontiers Laboratory, New York University Stern School, New York, New York, USA}

\affil[9]{MIT Media Lab, Massachusetts Institute of Technology, 75 Amherst St, Cambridge, MA 02139, USA}

\date{}

\begin{document}

\maketitle

\begin{abstract}
Machine olfaction is rapidly emerging as a transformative capability, with applications spanning non-invasive medical diagnostics, industrial monitoring, agriculture, and security and defense. Recent advances in stabilizing mammalian olfactory receptors and integrating them into biophotonic and bioelectronic systems have enabled detection at near single-molecule resolution thus placing machines on par with trained detection dogs. As this technology converges with multimodal AI and distributed sensor networks imbued with embedded AI, it introduces a new, biochemical layer to a sensing ecosystem currently dominated by machine vision and audition. 

This review and industry roadmap surveys the scientific foundations, technological frontiers, and strategic applications of machine olfaction making the case that we are currently witnessing the rise of a new industry that brings with it a global chemosensory infrastructure. We cover exemplary industrial, military and consumer applications and address some of the ethical and legal concerns arising. We find that machine olfaction is poised to bring forth a planet-wide molecular awareness tech layer with the potential of spawning vast emerging markets in health, security, and environmental sensing via scent. 
\end{abstract}

\section{Introduction}
Throughout evolution, life has depended on exchanging chemical messages through the air and water. Olfaction, the sense of smell, enables organisms to detect and act upon minute traces of volatile compounds, guiding them toward food and mates, and even giving them the ability to detect human disease by sniffing \citep{DeWeerdt2024}. While machine vision and audition now rival the human eye and ear, our technologies have long struggled to match the nose \citep{Covington2021}. At its core, biological olfaction is a feat of molecular matchmaking and pattern recognition: small molecules called odorants (typically under 350 atomic weight) bind with high specificity to much larger membrane-bound receptors (typically ~30,000 molecular weight), triggering even whole-organism behavioral responses to vanishingly small chemical cues. Traditional analytical systems such as Gas Chromatography - Mass Spectrometry (GC-MS), Liquid Chromatography-Mass Spectrometry(LC-MS), and odor fingerprinting approaches involving a variety of sensors have often been attempted as noses. These have traditionally used metal oxide sensors, and interferometric, surface-plasmon resonance (SPR), quartz crystal microbalance (QCM) or THz-time domain spectroscopy (THz-TDS) platforms \citep{laplatine, Songkhla2025QCM, Chapalo2024, Qin2023, Koch2023THz}  and have long achieved remarkable sensitivities under laboratory conditions,  but none have led to a machine olfactor with not the contextual nuance, specificity, generalizeability, or speed required to truly mimic the capabilities of the biological sense of smell and the nervous systems that use it.

Two key barriers stood in the way of machine olfaction as it was poised to join the club of machine vision and audition: 1) the ``limit of detection'' as machine olfactors could not reliably respond to single-molecule events and 2) the ``limit of recognition'', the ability to decode complex scent characters in plumes mixed in dynamically fluctuating background noise. Both are being overcome. A breakthrough came with the successful integration of stabilized mammalian olfactory receptors into photonic and electronic devices \citep{kim2009, goldsmith2011, yoo2021, kim2023}. These systems, when married with modern AI, including deep neural networks and large language models \citep{sundararajan2025enhancing,boscher2024sense,feng2025smellnet,liang2024foundations} are showing promise in real-world scent recognition.

Programs such as DARPA’s Dog Nose and Real Nose \citep{Weinberger2007,Harrison2019,DefenseDaily_RealNose2008} catalyzed this shift, demonstrating that embedding biological receptors into devices was powerful. Recent developments now promise to allow biomachine noses to detect disease signatures in urine, sniff out chemical threats in the field, and discriminate among thousands of scents conferred by wafting odorants with speed and accuracy \citep{Maurer2022,Rotteveel2025,Yoo2022,DeWeerdt2024}.

As these systems mature, receptor-based sensors, sensor arrays, and AI classifiers are converging into deployable platforms. Peer-reviewed research confirms single-molecule detection in engineered membranes, robust olfactory classification, and application across healthcare, agriculture, and defense \citep{guest2021COVID,warli2023}. As a result, machine olfaction is rapidly emerging from neglected to a foundational layer of intelligent sensing, poised to complement and in some cases outperform light and sound sensing in next-generation sensors, AI and robotics.

\section{Emerging Technological Frontiers}

Modern olfactory sensors draw inspiration from the mammalian nose, using materials such as graphene, metal oxides, and organic films in sensor arrays that detect a wide range of volatile organic compounds. These arrays, paired with pattern recognition and AI, can achieve remarkable sensitivity, as individual receptors are capable of triggering upon capturing single molecules. Some advanced platforms integrate microfluidics to better deliver and filter samples, while embedded AI and progress in sensor-actuator loops is enabling local processing and autonomous adjustments, akin to nostril-like sniffing in response to odor plumes. This design follows embodied cognition, the idea that the body itself contributes to perception and decision-making, not just a top-down brain commander. The act of sniffing is often adjusted reflexively by snout, nostril and lung muscles based on environmental cues, without need for conscious control \citep{Wachowiak2011}. Similarly, embedded systems with local sensors and actuators can respond in real-time, without  a centralized command structure. This ``sense-to-act'' loop improves adaptability, especially in dynamic environments. The olfactory system exemplifies this concept: it performs early processing in the nose and olfactory bulb, enabling rapid, efficient responses \citep{johnson2003}. Mimicking this decentralized and embodied intelligence allows for more robust AI systems, whether in robots, drones, or as part of large-scale sensor networks. These systems can operate locally yet remain connected to central nodes, supporting distributed intelligence that scales across platforms and environments~\citep{liang2020think}. While mimicking the embodied intelligence of olfactory systems enables remarkable adaptability, it also brings ethical considerations. The capacity to detect the presence of sensitive compounds or tell-tale scent signatures left behind by individuals unaware of possible detection introduces questions around privacy and surveillance, while decentralized intelligence complicates accountability and decision-making. As these systems expand into healthcare, environmental monitoring, and security, their design must balance technical performance with transparency, responsible use, and equitable access.

\subsection{Machine Learning-Driven Odor Recognition}
Machine learning methods, including deep convolutional neural networks and decision forests, are increasingly used to interpret high-dimensional sensor data~\citep{brattoli2011odour,sung2024data,yan2015electronic}. These models identify latent features in chemical signatures and classify them with accuracy exceeding 90\% in specific diagnostic and industrial applications. A critical issue is the presence of bias in olfactory datasets, often due to limited representation of environmental or demographic variability. The need for debiasing is well illustrated in recent research using GC-MS data to detect prostate cancer, where machine learning models initially overfitted to cohort-specific background odors rather than disease-relevant markers \citep{Rotteveel2025}. This challenge echoes broader AI bias issues documented in medical imaging AI models \citep{Obermeyer2019}.

\subsection{Multisensory Artificial Intelligence}

Multisensory artificial intelligence is a vibrant multidisciplinary research field that designs computer agents that can perceive, reason, and interact through multiple communicative modalities, including linguistic, acoustic, visual, tactile, sensory, and physiological messages \citep{liang2024foundations,liang2023quantifying}. Multimodal AI systems can be of great impact in diverse scientific areas bringing practical benefits, such as supporting human health and well-being~\citep{acosta2022multimodal,dai2025qoq}, enabling multimedia content processing~\citep{liang2021multibench,liang2024hemm}, and enhancing real-world autonomous agents~\citep{durante2024agent,xie2024large}. Multisensory models trained on combined vision, audio, and olfactory data have already demonstrated superior context interpretation in many medical sensing fields \citep{dai2025climb,liang2024foundations}. This fusion aligns with a broader AI trend toward embodied cognition and contextual awareness~\citep{duan2022survey}. By leveraging advanced multimodal AI techniques, there is potential to combine olfaction with other human senses to improve resilience and inference capability in real-world applications.
One example of such a technique is called cross-modal translation \citep{liang2024foundations,pham2019found}, where multimodal models can take in one modality and generate another, such as using a visual image to predict how the substances (whether objects such as food items or odorant molecule identity and concentration descriptors) in the image would feel, taste and/or smell to a human~\citep{brewster2006olfoto,rombach2022high}. These cross-modal translation models could be trained on paired modalities (e.g., image and its corresponding smell descriptors) to decode chemical or sensor representations of smell. Another technique is called cross-modal fusion \citep{liang2024foundations}, which investigates whether jointly learning from images, audio, and scent can improve performance on individual modalities such as visual object recognition, speech recognition, and smell recognition~\citep{huh2024position,radford2021learning,tjandrasuwita2025understanding}. Prior work has shown that jointly learning from human verbal and nonverbal communication can improve performance as compared to learning from only verbal spoken words \citep{ong2025human,zadeh2020foundations}. Similarly, today's large language models are increasingly trained with visual and auditory inputs \citep{li2024multimodal}. Studying how olfaction can be learned jointly with visual and language modalities can lead to learning algorithms that learn like humans do, grounded in multisensory inputs~\citep{bisk2020experience,liang2022brainish}.
In addition to multimodal AI models, a key driver for this line of research is multisensory scent environments, where data associate with the same physical space and processes ongoing are collected from multiple sensing modalities \citep{feng2025smellnet}. For example, collecting unified datasets of foods and beverages with smell sensor readings, photos, text descriptions, chemical molecular analysis \textit{etc.} merged into generalized ``Synesthetic Memory Objects''. These rich multimodal datasets are essential for large-scale multimodal models \citep{antol2015vqa,li2024multimodal,liang2022high}, and have contributed significantly to today's multimodal foundation models.

\subsection{Non-Invasive Medical Screening and Diagnostics}
While electronic noses (``e-noses'') have been around for decades they are now being supplemented with new ``biomachine olfaction'' technologies. These can use signatures created by their primary sensing layers' interaction with volatile organic compounts (VOCs) found in human emissions and have been emerging as part of a growing suite of promising tools for the non-invasive diagnosis of diseases. Such systems offer a potentially disruptive new capacity for earlier detection and improved patient outcomes through rapid, cost-effective screening methods for detection of  neurodegenerative, infectious, and oncological diseases. For instance, in the case of Parkinson's disease an AI-driven e-nose analyzing sebum samples achieved a sensitivity of 91.7\%, highlighting the role of skin-emitted scent as a generalized diagnostic biomarker \citep{american_chemical_society2022}. Notably, Parkinson's has also been shown to be diagnoseable by Ms Joy Milne, a famous super-smeller nurse who can diagnose Parkinson's by smelling patients' t-shirts \citep{kwon}. Similarly, advances in biosensor engineering have yielded electrochemical sensors capable of detecting amyloid-beta oligomers which are key pathological indicators of Alzheimer's disease with high sensitivity and specificity \citep{park}. Such technologies have also shown promise as infectious disease diagnostics e.g. the eNose-TB project in Indonesia evaluated on active tuberculosis screening using breath analysis, achieved a sensitivity of 95\% and specificity of 82\%, underscoring the technology's relevance in resource-limited healthcare settings \citep{center2018}. In oncology, systematic reviews support the role of electronic noses in the non-invasive detection of multiple cancers through exhaled breath analysis, affirming the feasibility of this approach, citing its diagnostic performance across several cancer types \citep{scheepers} including notably prostate cancer. Urine-based detection, using trained dogs sniffing out prostate cancer in urine can reach up to 99\% sensitivity \citep{taverna2015}. Dogs can even detect COVID-19 by smelling sweat samples collected noninvasively from the skin, achieving up to 98\% sensitivity and 100\% specificity, even in asymptomatic individuals \citep{Maurer2022}. RealNose.ai reports here that it had to start with urine in their lab-to-market transition because urine collection is already considered a well-established medical route to diagnosis via the presence of a myriad of potential biomarkers dissolved in urine. RealNose.ai also has positioned themselves to anticipate other-than-urine modalities that will undoubtedly eventually overtake starting with those that require even less effort (e.g., skin (sweat), breath). Again these capabilities raise questions about privacy, consent, and the handling of intimate biological information. Non-invasive olfactory diagnostics could enable earlier detection, but they also risk stigmatization or discrimination if results are inaccurate, leaked or misused. Ensuring accuracy, patient autonomy, data protection, and transparency in how scent character biomarkers are analyzed and interpreted will be critical for trust and equitable adoption, and legal frameworks currently applied to image and sound data would have to be adapted to scent.

\section{Emerging Computational Frontiers}
\subsection{Quantum Algorithms for Chemical Pattern Recognition}
Quantum-enhanced classifiers offer potential speedups in odorant-receptor matching and molecular similarity detection. Simulating olfactory binding events using quantum Hamiltonians could support more effective training of machine learning models for olfaction. Inspired by physical principles, quantum algorithms organize high-dimensional VOC data in novel ways, sometimes revealing patterns that classical methods miss. This can help reduce complex olfactory data into more separable subspaces, improving classification performance and insight into scent-based interactions. Recent re-ignition of the conversation about using biomatter as an ambient temperature, noise-resilient substrate for quantum computing  might eventually converge as proposals for using biological structures that are affected by small molecules are emerging \citep{NEMAMDVN2025}.

\subsection{Neuromorphic Architectures for Edge AI}
Olfactory signals are sparse and temporal—well-suited to neuromorphic chips running spiking neural networks. These architectures process sensory data with ultra-low power, enabling real-time chemical recognition on drones, wearables, or remote sensor nodes \citep{LeBow2021}. This edge-based approach reduces latency, preserves privacy, and ensures robustness in disconnected environments. RealNose mimics neurobiology as it is a platform made of three parts that work together similar to how the mammalian nose-brain-memory system works using actual, biological receptors, compact bioelectronics and or photonics to do the measurement, and software that recognizes patterns. Instead of chasing a list of single molecule names and concentrations, the system reads the overall scent signature of a sample and learns to distinguish foreground from backrground, i.e. context awareness is key to olfaction and embedded AI is a big part of the enabling idea here. The same core design runs in a lab instrument for healthcare and in a mobile unit for defense upon ruggedization ad changes in the user interface. RealNose also has ambitions beyond building the necessary hardware, it is also in the initial stages of building a growing library of labeled scent signatures from clinical specimens and controlled reference mixtures. Every record carries basic context like when, where, and how it was collected plus a widely varying donor's medical history. Models are retrained as the library grows, and each release is versioned so results can be traced during audits, clinical studies, or field investigations and allows RealNose to expect improvements in the clustering accuracy as the dataset grows and is continuously re-evaluated, just as a nose learns to refine its recognition capabilities with repeated use. The device uses short inhale and purge cycles to improve capture at very low levels and in changing air and multiple sniffs are used, again mimicking biology and its near universal sniffing rate. Recent comparative work demonstrates that sniffing frequencies remain confined to a narrow range across mammals, from rodents to elephants, clustering around 4--8\,Hz despite orders of magnitude differences in body size \citep{rygg2021sniffing,schneider2023nasal}. This relative constancy suggests that the ~5\,Hz rhythm is not dictated purely by fluid mechanical scaling, but reflects the temporal tuning of olfactory receptor and neural circuit architectures that rely on the sniff cycle as a fundamental processing frame \citep{shusterman2011sniffing,grimaud2023olfactory}.

\section{Forecasting the Proliferation of Machine Olfaction}

The trajectory of olfactory sensor development mirrors earlier sensor revolutions in cameras and microphones. These were once exclusive to scientific instruments, yet are now essential parts of  billions of mobile and embedded systems and devices worldwide. They enable AI-powered medical tasks such as cervical cancer screening via smartphone imaging. Sound-of-cough-based disease classification and self-administered ultrasound diagnostics such as the handheld ultrasound Butterfly iQ are growing with several clinical evaluations now benchmarking performance against devices in point-of-care ultrasound settings showing comparable image quality, usability, and diagnostic concordance across modalities \citep{PerezSanchez2024}. Barometers, added to a smartphone for the first time in 2011 \citep{ye}, were repurposed beyond GPS calibration to detect building altitude shifts and enclosed space movement signatures \citep{Han2025, Monteiro2016,LSP2014Baro}. These examples suggest that new sensor technology often unlocks applications unforeseen by the original designs, a case of technology evolution bearing a striking analogy to biological evolution and the phenomenon of \textit{exaptation} \footnote{\textit{Adaptation}, the well-known mechanism of incremental improvement, allows an organism to better compete within its environment (the proverbial longer beak, or classic example of more effective camouflage). But evolution also allows for a more abrupt type of transition: \textit{exaptation}, evolution's version of ``off-label use''. Exaptation is when an existing feature takes on new, unintended function(s) such as for instance fingers meant to grasp objects being used for sign language. Where adaptation is the steady workhorse, exaptation is a lateral leap that redefines the actual purpose of an existent system \citep{RoseMershin}.} in evolutionary biology. All this supports the argument for accelerated olfactory deployment due to known and yet-to-be-discovered synergistic (synesthetic) effectiveness multipliers resulting from deployment and use in the real world. 

Biomachine olfaction is however critically different to cameras, microphones, and barometers in the fact that it can reveal intimate personal biochemical signatures. As such, its development and deployment must be done with keen attention to the moral and legal ramifications, accompanied by appropriate ethical safeguards to protect human wellbeing.

The global market for olfactory sensors is in the early phases of significant expansion, driven by exceptional growing interest in non-invasive diagnostics, real-time environmental monitoring, and precision agriculture  now more urgently motivated by the globally-expanding spending on defense growing at a compound annual rate of 26.7\% \citep{ref1}.

Building on recent momentum, Ainos (NASDAQ: AIMD) has signed a five-year distribution deal with Solomon Technology Corporation to bring its AI Nose platform to industrial sectors across Asia. The partnership combines Ainos’ Smell Language Model (SLM) with Solomon’s Visual Language Model (VLM), creating a multisensory AI platform aimed at industries like semiconductors, petrochemicals, autonomous robotics, and healthcare. Given that Asia accounts for over 70\% of global semiconductor production and nearly half of all electronics manufacturing, the region offers a natural launchpad for widescale adoption. Pilot programs are expected to roll out in the second half of 2025, with commercial deployments targeted for 2026. This collaboration highlights how strategic partnerships are fast-tracking Commercialization and expanding market potential are gathering pace. The global electronic nose market, worth about \$29.8 billion in 2025, is projected to reach \$76.5 billion by 2032, an increase equivalent to compound annual growth of roughly 14\% or more than two and a half times its current size over the next seven years. Such a trajectory, while hardly frothy by technology standards, signals substantial upside for olfactory AI in industrial automation \citep{stocktitan2025}.

As this market accelerates, questions of data ownership and control over olfactory signatures become increasingly important, since scent data, potentially much more so than visual or acoustic signals, can reveal deeply personal health information \citep{demir2023}.

\subsection{Human and Animal Healthcare Diagnostics and Screening}

Olfactory sensors show increasing promise in clinical diagnostics and screening. Recent studies have validated their use for detecting respiratory diseases such as COVID-19 by detecting signatures conferred by volatile organic compounds in exhaled breath, demonstrating diagnostic accuracy comparable to PCR testing in controlled settings \citep{ref2}. The success of detection dogs, such as those trained under the Medical Detection Dogs UK COVID Canine Detection program, \citep{guest2021COVID} has provided biological benchmarks for machine olfaction systems. Precision agriculture systems increasingly incorporate detection of volatile organic compounds (VOCs), enabling assessment of crop stress and optimization of irrigation, pesticide use, and harvest timing \citep{kim}. In livestock, olfactory sensors can detect disease early and track reproductive health \citep{neethirajan}. Olfactory sensors embedded in breathalyzers and smart masks are already under development \citep{heng2024}, smartphones are likely to follow. We note that planet-wide whole-city or even whole-country shutdowns would be unimaginable during the recent COVID-19 pandemic if smartphones were equipped with COVID detectors as capable as trained dogs in detecting their owner's infection status with literally one sniff \citep{guest2021COVID}.

\subsection{Environmental and Air Quality Monitoring}
Urban deployments of e-noses could allow for high-resolution mapping of pollutants like ozone, ammonia, and benzene \citep{temkar}. Integrated into smart city platforms and available to geolocated smartphones \citep{suriano}, real-time data for public health planning and emergency response could become a powerfully disruptive predictive capability akin to predicting the landfall of a hurricane with precision only for pollution, infestation, disease and other preventable large-scale disasters.
Scent patterns collected by olfactory- gas- and chemo- sensing geolocated networks coupled with weather and satellite imagery data showing blooms of plants and algae can be integrated to study air polution and other trends, triangulate leakage source points taking into account the complex ways that scent moves in open and enclosed spaces in chemical and industrial installations, and can be of significant help in identifying and backtracking the sources of illegal polution and dumping \citep{mccaul}. Integration with Internet of Things (IoT) infrastructure would allow for real-time alerts and predictive modeling via distributed sensor networks~\citep{baris2025foundation,mo2024iot}.
Here, too, olfactory environmental monitoring introduces ethical questions about surveillance and ownership of such potentially lucrative and / or disruptive data. Communities may benefit from early warnings about air quality hazards, but they also deserve a voice in how sensor networks are deployed, who controls the resulting data, and whether the insights are used solely for public good or also for commercial advantage, and to whom these advantages might be conferred.

\subsection{Food, Beverage and Medicine Quality Control}

In industrial and small-scale food growing and processing, machine olfaction enables rapid spoilage detection and quality assurance, offering cost-effectiveness and addressing waste and spoilage associated losses. E-noses have been successfully applied in meat, dairy, and beverage production lines \citep{ref4} and have been evaluated for online-monitoring of biopharmaceutical processes by GlaxoSmithKline, MIT and RealNose.ai.

\subsection{Security, Defense and CBRN}

Chemical and radiological agents often alter the microflora, plant volatiles, and soil chemistry of contaminated environments. These changes emit unique odor profiles that can be detected even after the source is removed. For example, exposure to nerve agents like Sarin can cause persistent VOC changes in flora, while radiation alters microbial metabolism and emissions. Distributed olfactory systems could serve as silent sentinels for CBRN event detection \citep{FrontiersVOCs2023,HortRes2023_VOCsStress,FEMS2024_ExtremeMVOCs,Microorganisms2023_mVOCs,PCE2024_IonizingRadiationPlants,BMCPlantBiol2025_UV,SciRep2025_CWA_Simulant,Sensors2025_MOS_CWA,SPIE2010_PortableENoseCWA}. Deploying olfactory systems for defense and CBRN detection inevitably intersects with global security and arms control debates. These tools could provide early warning and deterrence, but they also risk escalation if used for covert monitoring or in contested territories. Ethical deployment must therefore balance national security interests with international norms, emphasizing transparency and adherence to treaties\citep{JAgEnvEthics2019_BiosurvEthics}.The sensitivity of machine olfaction to trace chemical compounds enables applications in explosive detection and narcotics/contraband screening by deployment in ports and airports as passive surveillance tools \citep{ref6}. Beyond point sensors, \emph{platform} approaches such as where dual-use startups reruggedize and re-deploy interchangeable bio/chemosensor arrays originally meant for medical machine olfaction to the battlefield. A platform capable of being re-trained on anything with a scent profile expands the performance envelope and mission flexibility. Biomimetic “active sniffing” designs have already shown that artificial olfactory front-ends can exceed conventional continuous-suction samplers, improving trace vapor capture by an order of magnitude and boosting the detection performance of a commercial explosives detector in controlled tests \citep{Staymates2016}. Platform, trainable noses can be robot-mounted for standoff operations and ISR: recent open-access work in robotics demonstrates robust olfactory navigation and odor-source localization using sensor fusion (vision\,{+}\,olfaction) that outperforms single-modality baselines in obstacle-rich, turbulent environments—directly relevant to CBRN reconnaissance on quadruped “robot-dog” chassis \citep{hassan2024}. 

Olfactory navigation itself is recognized as an operationally useful capability: mammals, birds and insects exploit the temporal structure of turbulent odor plumes for mapping and wayfinding, a principle now informing the design of \emph{robotic noses} for search, patrol, and hazard approach/avoid behaviors \citep{PLOSBioNav2024}. In military/law enforcement settings, open-access reviews outline current and near-term uses of olfaction and electronic noses for explosives, force protection, and training—situating chemical sensing squarely within modern ISR and mission support \citep{MMR2017}. At the network level, DARPA’s SIGMA+ program explicitly targets persistent city-/region-scale CBRNE sensing through advanced detectors and analytics—an architecture well-matched to startup “platform” models that can field swappable payloads across fixed, mobile, and autonomous nodes \citep{DARPA_SIGMAPlus}. 

Critically, while detection dogs remain a strong benchmark, peer-reviewed evidence shows canine sensitivity and reliability are \emph{environment-dependent} (temperature/humidity), with measurable threshold degradations across energetic materials under operationally relevant extremes \citep{PLOSONE2024}. Platform e-nose systems engineered for adverse environments, augmented by biomimetic sampling and plume-aware navigation, offer resilient coverage and repeatability and can be coupled to quadruped unmanned ground vehicles for remote sampling and interdiction. Such systems are currently being evaluated or piloted for deployment in ports and airports as passive or semi-autonomous surveillance tools and as adjuncts to screening workflows \citep{Staymates2016}.

Robotic Navigation and Search-and-Rescue are clear cases where olfactory capability would enables drones and ground robots to locate victims in disaster zones by detecting human-derived VOCs (e.g., isoprene, acetone) and human-caused disturbances in microflora and fauna and other environmental changes indicative of human presence that can leave behind a learnable trace scent -specific to each location. These tools complement visual or infrared sensing, especially useful in low-visibility or GPS-denied environments. While life-saving potential is clear, ethical concerns emerge around dual-use risks, particularly if olfactory navigation technologies are repurposed for surveillance or law enforcement beyond their original intent. Establishing governance frameworks that prioritize humanitarian use while preventing misuse will be vital to maintain public trust.

\section{Business and Market Potential}
The sensor market is currently projected to undergo significant development. According to Precedence Research, purely chemical sensors' market will increase from \$24.94 billion in 2024 to \$55.36 billion by 2034 \citep{precedenceresearch2024}, at a compound annual growth rate of 8.3\%. This robust growth is thanks to the expansion of the IoT. A notable breakthrough is the development of miniaturized, high-speed \citep{Dennler2024HighSpeed} electronic noses, capable of classifying odor and at least two startups using nature's own scent receptors: mammalian G-protein coupled receptors (GPCRs) \citep{RealNose2025} and insect ORCOs \citep{scentianbio}. 

As these and similar platforms scale, they increasingly generate data which can be analyzed and acted upon in real time. In parallel, researchers are advancing hybrid neuromorphic-Bayesian frameworks that integrate low-power spiking neural networks with probabilistic inference, offering adaptive and energy-efficient odor recognition well-suited for edge applications \citep{kausar2024efficient}. At the systems level, olfactory inertial odometry (OIO) integrates olfactory sensing with motion data, supplementing inertial navigation assisting robots' movement with scentmaps \citep{france2025olfactory}. Data limitations are also being addressed through diffusion model-based molecular augmentation, which expands the odorant chemical space for training and validation \citep{france2025diffusion}. 

On the materials front, transition-metal phosphides (TMPs) and nanostructured electrocatalysts are being engineered into advanced chemosensors with heightened sensitivity, stability, and selectivity, particularly in food safety and spoilage detection \citep{wei2024recent}. Complementary work focuses on miniaturization and cost reduction, yielding portable platforms for agriculture, environmental monitoring, and consumer application \citep{chemosensors2025}. In parallel, AI-driven frameworks are being applied to olfaction not only for odor classification, but also for gustatory-olfactory fusion, adaptive learning under sensor drift, and even personalized scent design \citep{aigustation}. 

According to McKinsey \& Company, the total projected economic value of IoT is \$3.9–\$11.1 trillion per year by 2025 \citep{mckinsey}. Much of this is driven by nascent sensor-enabled applications like healthcare, automation, environmental monitoring, and consumer electronics. Machine olfaction in particular opens up new layers of data, enabling the detection of signatures conferred by volatile organic compounds and other trace chemicals that are missed by electromagnetic and acoustic sensors.
As machine olfaction technology matures, standardization becomes critical for adoption. Standardization is key for  efficiency, reliability, safety, and interoperability across various use cases. The Institute of Electrical and Electronic Engineers (IEEE) P2520 focuses on developing consistent and unified terminology, performance benchmarks, and safety criteria \citep{danesh}. This creates a basis for widespread adoption accross established and new sectors such as consumer products and electronics.

Beyond industry, the Organisation for Economic Co-operation and Development (OECD) points to the use of such sensors in public services. Governments view sensors as the enabler of digital infrastructure, with high-profile applications such as improvements in urban planning, healthcare delivery, and disaster response \citep{oecd}. The OECD also asks for cross-sector data sharing, further maximizing public benefit. Specific applications range from real-time air quality alerts and elder care monitoring to the integration of olfactory sensors into national and environmental health systems.

\paragraph{Investment Implications}
As sensor technologies mature, machine olfaction is emerging from the laboratory into commercial deployment. For early investors, the appeal lies not in the clear trajectory toward an emergent new industry where first mover advantage can be crucial to growth. Platforms such as RealNose.ai are positioned to complement vision and acoustic arrays with an entirely new biochemical layer of intelligence, supplying real-time, unambiguous data streams as robust as satellite weather feeds. The potential return on capital is underpinned by demand in high-margin markets—medical diagnostics, defense, and industrial security—where decision-making advantages are valued at a premium. Interest from biobanks and collaboration with leading academic institutions such as the MIT Media Lab,  MIT Lincoln Laboratories, participation in CBRN military exercises and commitments by Cleveland Clinic and the International Agency for Research on Cancer (IARC) to provide samples (amongst a growing number of volunteer and commercial partners) and validation materials to RealNose.ai, demonstrate that this capability is advancing beyond concept into reproducible, clinically impactful use.  In the UK, NIHR‑linked sponsors Imperial College London, HealthTech Research Centre in In Vitro Diagnostics, University of Exeter with ExeCTU/PenTAG, Newcastle North East Innovation Lab, Clatterbridge Cancer Centre, and NIHR BioResource have issued Expressions of Interest to help the effort towards validation of machine olfaction solutions and NIHR funding routes, potentially opening up  trial infrastructure, recruitment at scale, and, most importantly for the training, biobanked cohorts.
 A distinct advantage of machine olfaction over analytical technigques is its ability to register residual signatures of people, pathogens, and toxic agents, without having to know upfront biomarkers in the form of lists of constituent molecules, instead relying on the overall scent character that can often linker long after its direct causes have dissipated. For investors, this represents not just incremental innovation but the opening of a new frontier in sensing, a new global industry  destined to generate enormous proprietary data streams and durable value in industries where early insight is decisive \citep{MassChallenge2025,RealNoseWebsite}.

\section{Ethical, Legal, and Regulatory Considerations}
We have touched upon the ethical, moral and legal concerns throughout the text and here we elaborate on how the widely distributed adoption of machine olfaction raises novel privacy concerns. Ambient VOCs may inadvertently reveal sensitive data such as substance use, metabolic, neurodegenerative,  chronic or infectious disease state or even pregnancy or parenthood or genetic relative status in individuals who do not consent to such information being publicly known about them. Regulatory bodies must address the legality of such passive sensing in public and private domains. Guidelines must define permissible use, retention, and interpretation of olfactory data, akin to HIPAA protections in health data and GDPR standards in digital privacy. \textit{The Osmocosm Foundation} is an MIT-launched, Boston-based Massachusetts  501(c)(3) non-profit public benefit foundation \citep{OsmocosmFoundation} has organized three machine olfaction-themed conferences where moot court competitions took place examining aspects of privacy, liability, mining  and ownership of scent data collected and interpreted with or without the consent of an individual.

An additional ethical dimension arises when machine olfaction is considered for security applications. Beyond privacy, the potential for misuse of such a technology comes with great ethical and humanitarian risks, such as the (im)precise targeting of individuals based on their biological signatures by rogue military forces. The possibility that this technology falls into the wrong hands, such as criminal regimes or non-state actors, should be taken into account before this technology is integrated into military environments. Through these parties, machine olfaction can be used for covert, repressive surveillance, terror operations, or targeted assassinations.

Developers of this technology should prioritize its responsible use, especially with regard to privacy and the protection of non-combatants. Maintaining transparency and collaborating with trustworthy partners will contribute significantly to preventing misuse. 

We recognize that it is also critically important that regulatory frameworks around machine olfaction that align with international humanitarian laws, such as the Geneva Conventions and the NATO's objectives (see the North Atlantic Treaty and NATO's Policy on the Protection of Civilians), are established. By focusing on oversight, transparency, and accountability in the integration of machine olfaction in military spaces, misuse can be prevented and ethical standards, as defined by democratic values and human rights, can be upheld.


\section{Conclusions}
Machine olfaction is moving rapidly from curiosity to capability, and soon to necessity. Just as GPS and mobile connectivity became indispensable within a decade, artificial olfaction is  on course to embed itself across industries at a pace rarely seen in the history of new sensor technology adoption. The proliferation of sensors at the edge is ushering in an era of airborne biochemical awareness. In consumer markets, this shift is already visible: the vastness of just the Chinese market for air purifiers and pollution detectors has normalized the idea of continuous environmental monitoring. It is not difficult to imagine a near future in which the very air we breathe is continuously sensed, classified, and labeled for health and security in real time, integrated with augmented and mixed-reality platforms. 

Consumers have long sought deeper transparency into what enters the body as food, drink, or medicine. Smartphones have only accelerated this demand, turning once-novel capabilities such as machine vision and speech recognition into everyday tools. Artificial olfaction extends this trajectory, offering a biochemical dimension to sensing that surpasses the limits of human biology in both speed and scale of information capture. Whether applied to diagnosing disease, monitoring air and water supplies, ensuring food safety, or detecting CBRN threats, machine olfaction is set to become a critical layer of 21st-century digital infrastructure a disruptive and enabling technology for nearly every sector of modern life.

\section{Outlook}

In a future increasingly shaped by advanced machine olfaction and embedded chemosensory AI, we are entering a transformative era: sensing technologies and their networks are currently evolving into something resembling a planetary nervous system. This system is poised to even have its own forms and functions reminiscent of familiar ``neurotransmitters'', and can still be molecular in nature despite the vastness of scale and lack of actual neurons. The analogy to  distributed-intelligence organisms such as anthills or slime molds is apt: cities infused with molecular sensor networks could continuously monitor air quality, soil health, and the biochemical signals of humans, plants, and animals. Much like the mycelial webs that connect trees in forests, these networks would relay vital-to-all life information in real time, creating a symbiosis between technology and nature at the level of molecular intimacy, bringing a new layer of perception to our existent light and sound sensing. In such a world, humanity could learn to ``read'' the air itself: the canvas for the chemical signatures of health, distress, danger, and opportunity. Machines detecting subtle shifts in these signatures would provide early warnings of disease, pollution, or ecological stress long before they grew into crises. Such chemosensory AI would, similar to modern meteorology, begin offering predictive, proactive tools. Agriculture could be transformed as smart farms deploy embedded scent sensors to track crop health, soil conditions, and growth cycles, enabling precision delivery of water and nutrients while minimizing waste. Cities would be capable of dynamically adjusting ventilation, transport, and sanitation systems in response to olfactory feedback, fostering cleaner, healthier, and more efficient and sustainable living conditions. On the personal level, handheld and wearable devices equipped with chemosensory functions could allow individuals to track their own biochemical states, detect allergens or pathogens in real time, and log cumulative exposure, an olfactory equivalent of a radiation dosimeter, but attuned to the chemical dimension of daily life. The next stage of digital infrastructure will not only see and hear but also smell, anchoring intelligence in the same signals that life itself has used since the very beginning: molecules.

\section*{Acknowledgments}

The work of AM, NS, AR, MK, ZO, ZK is supported by RealNose.ai -itself thankful for their early and critical support to: George Kung of KungHo Fund,  Armando Ca$\hat{n}$as and Robert Jursich of Mavericus Fund as well as Glenn Krevlin, Nelson Wang and Dr. Howard Kivell. Realnose.ai is a grateful awardee of US government research funds, a winner of NATO DIANA funding and Mass Challenge support for machine olfaction dual use development and a graduate of the New York University Stern School of Business’ Endless Frontiers Laboratory (EFL 2024-2025 Cohort) and has received excellent mentorship from the MIT Venture Mentorship Service and Louis Goldish. AM is grateful for the questions asked by participants in his MIT Sloan Executive Education class "Lab to Market the MIT Way" and the speakers, audience and other participants and contributors to the three annual Osmocosm.org conferences and moot court competitions focused on the rise of machine olfaction technologies. AM thanks urologist Clifford Gluck, MD for illuminating conversations focused on clinical relevance of machine olfaction in prostate cancer screening and early diagnosis and Despoina Gerasoudi for her support and expertise with sample handling and patient consent ethics. AM is thankful to Dimitri Ioannidis for crucial help with ideating the future legal aspects of olfaction technology development and Dr. Amelia Ruzzo and Dr. Shannon Johnson for helpful feedback on this manuscript. AM is thankful to Dr. Elcin Zan of the Cleveland Clinic and Dr. Zisis Kozlakidis of the International Agency of Research on Cancer for illuminating explanations of biobank processes necessary for medical machine olfaction training to FDA and CE standards. The RealNose.ai team are grateful to Dr. Tristan Rousselle and Dr. Cyril Herrier of aryballe for helping deploy biophotonic olfactory sensing. AM is thankful to Dr. Tzortzatou-Nanopoulou and the Digital Health Literacy and Policy Hub Foundation for the opportunities for outreach about machine olfaction including at the Lyon BioMed AI Conference 2025. We are grateful to Dr. Claire Guest and Medical Detection Dogs UK for spirited conversations about how machine and canine training can converge, Prof. Simmie Foster of the Harvard Laboratory for Hot and Cool Research and Massachusetts General Hospital for ideation and editing, and Dr. Howard Soule of the Prostate Cancer Foundation for his support of RealNose.ai efforts towards cancer-prostate specific medical machine olfaction.

\section*{Glossary}

\begin{description}

\item[501(c)(3)] \textit{Nonprofit, Tax-Exempt Organization:} A designation under the United States Internal Revenue Code for charitable, educational, scientific, or public benefit organizations that are exempt from federal income tax and eligible to receive tax-deductible contributions. The Osmocosm Public Benefit Foundation is organized as a 501(c)(3), supporting research, education, and public engagement in the field of machine olfaction and sensory sciences.

  \item[AI] \textit{Artificial Intelligence:} Simulation of human-like learning, reasoning, and perception in machines.

  \item[CBRN] \textit{Chemical, Biological, Radiological, and Nuclear:} Classification for high-risk hazardous materials.

  \item[CNT-FET] \textit{Carbon Nanotube Field Effect Transistor:} Nanotech component used in sensitive detection platforms.

  \item[De-biasing] The process of identifying and correcting systemic biases in training datasets, particularly in medical AI, to improve diagnostic fairness and accuracy across diverse populations.

  \item[Embedded AI] AI that operates locally within devices or edge systems, enabling autonomous data processing without cloud dependence.

  \item[Exaptation] Evolutionary concept where a trait developed for one purpose is repurposed for another — often invoked in tech to describe unintended but transformative use cases.

  \item[GC-MS] \textit{Gas Chromatography–Mass Spectrometry:} A technique for identifying compounds in chemical mixtures.

  \item[GDPR] \textit{General Data Protection Regulation (2016/2018):} 
  The European Union regulation governing personal data protection and privacy. GDPR sets strict rules for data collection, storage, processing, and transfer, emphasizing individual consent, transparency, and the right to access, rectify, or erase personal information.

  \item[GPCR] \textit{G Protein-Coupled Receptor:} A large family of membrane-bound proteins involved in signal transduction across biological systems. In olfaction, GPCRs detect VOC molecules (odorants) and activate intracellular responses, playing a central role in mammalian smell. Evolving more than a billion years ago in unicellular eukaryotes, GPCRs are ancient seven-transmembrane receptors conserved across nearly all animal, insect and many fungal lineages, though absent in plants.

  \item[HIPAA] \textit{Health Insurance Portability and Accountability Act (1996):} 
  A United States federal law that establishes national standards for protecting sensitive patient health information. 
  HIPAA mandates privacy, security, and breach-notification requirements for healthcare providers, insurers, and their business associates handling electronic health data.

  \item[ISR] \textit{Intelligence, Surveillance, and Reconnaissance:} A set of integrated military and security capabilities that collect, process, and distribute timely information on adversaries, environments, and operational conditions. ISR activities combine sensing platforms, data fusion, and analytic tools to support situational awareness, decision-making, and mission effectiveness.

  \item[LC-MS] \textit{Liquid Chromatography–Mass Spectrometry:} Analytical technique for separating and identifying molecules in liquid samples.

  \item[LLM] \textit{Large Language Model:} AI models trained on vast text corpora capable of sophisticated language tasks.

  \item[Machine Audition] Artificial systems designed to process auditory information, including sound localization, speech recognition, and acoustic scene analysis.

  \item[Machine Olfaction] Engineering of systems that detect and interpret chemical signals (smells) using sensors and AI, mimicking or surpassing biological olfaction.

  \item[Machine Vision] Technology enabling machines to process and interpret visual data for recognition, classification, or navigation.

  \item[ML] \textit{Machine Learning:} A subset of AI focused on algorithms that improve performance through experience and data without being explicitly programmed.

  \item[Odorant] A volatile chemical compound capable of eliciting a sensory response via olfactory receptors.

  \item[OIO] \textit{Olfactory Inertial Odometry:} Navigation method combining olfactory sensing and motion tracking.

  \item[OR] \textit{Olfactory Receptor:} A subtype of GPCR found in olfactory sensory neurons, specialized for detecting specific odorant molecules. Each OR responds to a range of odorants, contributing to combinatorial encoding of smell.

  \item[ORCO] \textit{Olfactory Receptor Co-Receptor:} A highly conserved ion channel protein essential for olfaction in insects, ORCO partners with specific odorant receptors (ORs) to form functional complexes that transduce chemical signals into neural activity similarly to ORs in mammals that are GPCRs.

  \item[THz-TDS] Terahertz time-domain spectroscopy offers valuable insights into material properties, its reliance on long interaction optical paths to accumulate sufficient absorption events makes it impractical for trace volatile detection at parts-per-trillion levels, where the ultra-low photon absorption cross-section necessitates either very high optical power or ultra-long integration times. By contrast technologies leveraging nature's own evolved biomachinery to capture low-concentration molecules (GPCRs) are free of this problem.

  \item[Quantum Biology] Study of quantum effects (e.g., tunneling, coherence) in biological systems such as smell, photosynthesis, and neural function.

  \item[VOC] \textit{Volatile Organic Compound:} Organic chemicals that evaporate easily and contribute to detectable odors.
\end{description}

\bibliographystyle{unsrtnat}   

\bibliography{references}

\end{document}